\documentclass[%
 reprint,
superscriptaddress,
 amsmath,amssymb,
 aps,
prb,
]{revtex4-2}

\usepackage{graphicx}
\usepackage{dcolumn}
\usepackage{bm}
\usepackage[colorlinks,linkcolor=blue,anchorcolor=blue,citecolor=blue,urlcolor=blue]{hyperref}

\begin{document}

\title{Exotic quantum light-matter interactions in bilayer square lattices}

\author{Xing-Liang Dong}
\affiliation{Ministry of Education Key Laboratory for Nonequilibrium Synthesis and Modulation of Condensed Matter, Shaanxi Province Key Laboratory of Quantum Information and Quantum Optoelectronic Devices, School of Physics, Xi’an Jiaotong University, Xi’an 710049, China }
\affiliation{Theoretical Quantum Physics Laboratory, Cluster for Pioneering Research, RIKEN, Wakoshi, Saitama 351-0198, Japan}

\author{Peng-Bo Li}
\email{lipengbo@mail.xjtu.edu.cn}
\affiliation{Ministry of Education Key Laboratory for Nonequilibrium Synthesis and Modulation of Condensed Matter, Shaanxi Province Key Laboratory of Quantum Information and Quantum Optoelectronic Devices, School of Physics, Xi’an Jiaotong University, Xi’an 710049, China }

\author{Jia-Qiang Chen}
\affiliation{Ministry of Education Key Laboratory for Nonequilibrium Synthesis and Modulation of Condensed Matter, Shaanxi Province Key Laboratory of Quantum Information and Quantum Optoelectronic Devices, School of Physics, Xi’an Jiaotong University, Xi’an 710049, China }

\author{Fu-Li Li}
\affiliation{Ministry of Education Key Laboratory for Nonequilibrium Synthesis and Modulation of Condensed Matter, Shaanxi Province Key Laboratory of Quantum Information and Quantum Optoelectronic Devices, School of Physics, Xi’an Jiaotong University, Xi’an 710049, China }

\date{\today}

\author{Franco Nori}
\affiliation{Theoretical Quantum Physics Laboratory, Cluster for Pioneering Research, RIKEN, Wakoshi, Saitama 351-0198, Japan}
\affiliation{Center for Quantum Computing, RIKEN, Wakoshi, Saitama 351-0198, Japan}
\affiliation{Physics Department, The University of Michigan, Ann Arbor, Michigan 48109-1040, USA }

\begin{abstract}

We investigate quantum emitters (QEs) interacting with a photonic structured bath made of bilayer square lattices, where the resonance anti-crossing between the energy bands opens a symmetric middle energy gap.
Due to the intrinsic chiral symmetry of the bath and interactions with the square-like band-edges,
the QE-photon dressed states generated in this inner bandgap are \emph{odd-neighbor, robust, and anisotropic},
when the emitters' transition frequencies lie in the middle of the bandgap.
We also use giant artificial atoms to engineer and modify the dressed states' patterns.
Exotic bound states can lead to spin models with symmetry protection, resulting in fascinating many-body phases. As an example, we show that this proposal can be used to generate
both edge states and corner states in the generalized 2D Su-Schrieffer-Heeger (SSH) model.
This work opens up new avenues for research into innovative quantum many-body physics and quantum simulations with photonic or phononic multilayer structures.

\end{abstract}

\maketitle


\section{Introduction}
Bilayer two-dimensional (2D) materials offer a flexible platform for studying strongly correlated electron movements near electronic flat bands in moir\'e superlattices
\cite{ROZHKOV20161,Cao2018Correlated,Cao2018Unconventional,2dmaterials}. A rapidly emerging field is to examine optical and acoustic analogs of bilayer 2D materials
\cite{PhysRevLett.120.116802,PhysRevB.101.121103,PhysRevB.102.180304,PhysRevLett.126.136101,Gardezi_2021,
PhysRevLett.126.223601,PhysRevB.103.214311,Tang2021Modeling}.
In this field, numerous theoretical works have predicted the localization and condensation
of light in bilayer photonic crystals
\cite{PhysRevLett.126.223601,Wang2020Localization,Fu2020Optical}.
The photonic counterparts have a distinct advantage in terms of flexibility, which includes the ability to construct a variety of lattice shapes and tune the coupling strength between layers via interlayer separation or other methods \cite{PhysRevLett.126.136101,PhysRevLett.126.223601,PhysRevA.100.053604}.  Beyond fundamental research focused on photonics, it is also very appealing to investigate light-matter interactions with quantum emitters (QEs) in photonic layered structures \cite{Tang2021Modeling,Strongly2022Bloch,PhysRevLett.128.113601}.

Long-range tunable coherent dipole interactions, on the other hand, mediated by bound states (BSs) produced within the bandgap of structured photonic baths, have attracted considerable interest due to their possible applications in quantum simulations
\cite{PhysRevLett.64.2418,Quantum2015Douglas,Subwavelength2015AGT,
PhysRevA.93.033833,Hood10507,PhysRevResearch.2.043213}.
Floquet engineering spin-spin interactions by periodic driving \cite{PhysRevResearch.2.013121,PhysRevResearch.3.013025}, tailoring the geometry of photonic interfaces \cite{doi:10.1021/acsphotonics.8b01455,PhysRevResearch.2.023003,PhysRevLett.126.203601,PhysRevLett.128.013601,Roccati:22}, and exploiting  giant atoms with nonlocal coupling to several positions of the bath \cite{Gustafsson207,PhysRevLett.120.140404,PhysRevLett.122.203603,PhysRevA.102.013709,
Kannan2020Waveguide,PhysRevLett.126.043602,PhysRevA.106.033522,PhysRevA.104.053522,
Kockum2021,PhysRevA.90.013837,PhysRevA.107.013710,PhysRevResearch.6.043222} are some of the approaches proposed to develop more innovative spin models.
Remarkably, QEs interacting with topological photonic lattices produce exotic BSs that can be chiral, resilient, and exhibit power-law scaling \cite{PhysRevA.97.043831,Belloeaaw0297,PhysRevLett.125.163602,PhysRevX.11.011015,
PhysRevLett.126.063601,PhysRevA.103.033511,PRXQuantum.3.010336,Photonic2022Saxena,Luca2021Dressed,
PhysRevLett.129.223601,PhysRevA.106.053517}.
However, previous works mostly focus on single-layer 2D or multi-layer 3D photonic structures; in stark contrast, photonic bilayer  structures possess very distinct band shapes and new features that have yet to be explored for light-matter interactions.

In this work, we predict a new type of QE-photon dressed states and coherent interactions
between QEs in photonic bilayer square lattices.
By taking advantage of only band crossing in a bilayer structure,
we can open a middle energy gap around 2D Van-Hove singularities without breaking the chiral symmetry of the lattices.
The intrinsic chiral symmetry of the bath together with the square geometries of the band-edges
give rise to the emergence of odd-neighbor (the photon profile vanishes to the even-neighbor sites of the QE),
robust, and anisotropic BSs when tuning the QE's frequency to the Van-Hove points of the bare bands.
We also use giant atoms in 2D to consider quasi-1D patterns and purify the anisotropy of the odd-neighbor BSs,
which cannot be implemented by pointlike emitters.
These BSs can mediate chiral symmetric and anisotropic spin-spin interactions, which have wide applications ranging from robust entanglement distributions to the simulation of nontrivial spin models.
Through a specific spin array arrangement, we can simulate a generalized 2D SSH model
that can support both edge modes and corner-localized modes.

\section{Model and Hamiltonian}

\begin{figure}[t]
\includegraphics[scale=0.18]{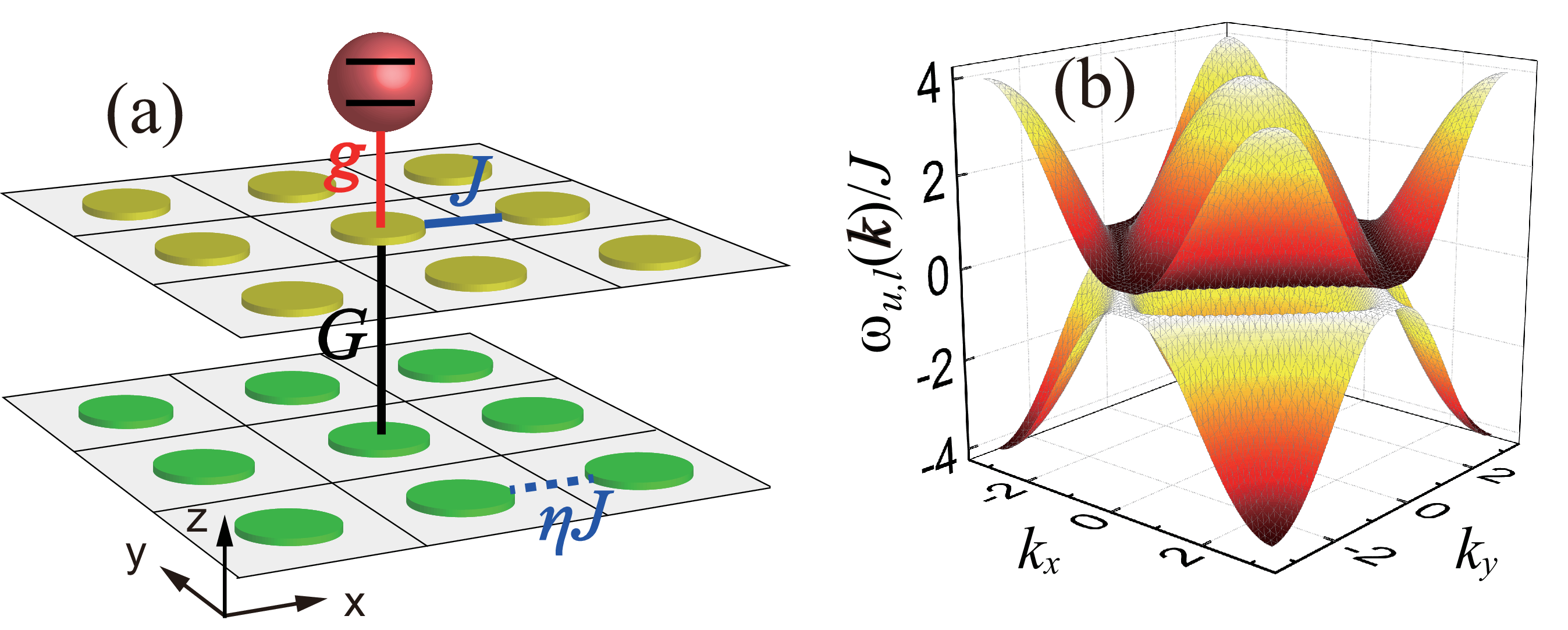}
\caption{\label{fig1}
(a) Schematics of emitters in bilayer square lattices.
The yellow and green disks represent lattice sites consisting of bosonic modes.
Here $J$ and $\eta J$ denote the intracouplings and $G$ is the intercoupling strength.
Emitters are coupled to the lattice sites with the interaction strength $g$.
(b) Corresponding energy band structure of the bath shown in (a),
with the middle bandgap $\varepsilon_\text{gap}=2G$.
Here, $\eta=-1$ and $G=J/4$.}
\end{figure}

As illustrated in Fig.~1(a), we consider a photonic double-layer structure,
where $J$ and $\eta J$ respectively describe the hopping amplitudes
between the nearest neighbour lattices in the top and bottom layers,
and $G$ denotes the interlayer coupling strength.
Here we mainly focus on the parameter regime $\eta<0$.
In this regime, we are particularly interested in the $\eta=-1$ case (without loss of generality)
and this bosonic Hamiltonian with anti-symmetric coupling corresponds to an anti-ferromagnetic coupling,
which is given by ($\hbar=1$)
\begin{eqnarray}\label{ME1}
\hat{H}_B&=&J\sum_{<\boldsymbol{n},\boldsymbol{m}>}(\hat{a}_{1\boldsymbol{m}}^\dag\hat{a}_{1\boldsymbol{n}}
-\hat{a}_{2\boldsymbol{m}}^\dag\hat{a}_{2\boldsymbol{n}})\notag\\
&+&G\sum_{\boldsymbol{n}}\hat{a}_{1\boldsymbol{n}}^\dag\hat{a}_{2\boldsymbol{n}}+\text{H.c.},
\end{eqnarray}
where $\hat{a}_{1\boldsymbol{n}}$ and $\hat{a}_{2\boldsymbol{n}}$ are annihilation operators of bosonic modes
at the position $\boldsymbol{n}=(n_x,n_y)$ in layer 1 and layer 2, respectively.
Here, we write the bath Hamiltonian in a frame rotating at the frequency $\omega_1\!=\!\omega_2$,
with $\omega_{1,2}$ being the cavity's resonance frequency in the two arrays.
We also ignore polarization and choose a scalar interaction
(we may assume the E-field vector in a direction perpendicular to the plane),
by further considering the coupled two-level systems, as discussed below.
We will study the effect of polarization in future work.
Note that this bath is equal to the one with $\eta=+1$ and alternating phases $(-1)^{n_x+n_y}$
in the interlayer couplings \cite{PhysRevA.95.013812}, where a photon with momentum $\boldsymbol{k}$
in the top layer is coupled to a photon with momentum $\boldsymbol{k}-(\pi,\pi)$ in the bottom layer.
One can easily check that the energy band with such a momentum shift is the same as the one with negative hopping rates.
In fact, both of them correspond to the introduction of a uniform effective flux of $\pi$ in each plaquette
$\{\hat{a}_{1\boldsymbol{n}},\hat{a}_{2\boldsymbol{n}},\hat{a}_{1\boldsymbol{n}_p},\hat{a}_{2\boldsymbol{n}_p}\}$,
with $\boldsymbol{n}_p-\boldsymbol{n}=(0,\pm1)$ or $(\pm1,0)$.
We next transform the Hamiltonian into $\boldsymbol{k}$ space by introducing the Fourier transformation
$\hat{a}_{\boldsymbol{k}}=1/\sqrt{N}\sum_{\boldsymbol{n}}e^{-i\boldsymbol{k}\cdot\boldsymbol{n}}\hat{a}_{\boldsymbol{n}}$,
with $\boldsymbol{k}$ being the wave vector (for convenience, we assign the lattice constant $d_0=1$ here).
In terms of $\hat{V}^\dag_{\boldsymbol{k}}=(\hat{a}^\dag_{1\boldsymbol{k}},\hat{a}^\dag_{2\boldsymbol{k}})$,
the bath Hamiltonian in momentum space can be written as
$\hat{H}_B=\sum_{\boldsymbol{k}}\hat{V}^\dag_{\boldsymbol{k}}\hat{H}(\boldsymbol{k})\hat{V}_{\boldsymbol{k}}$,
with the kernel
\begin{eqnarray}\label{ME2}
\hat{H}(\boldsymbol{k})={\left(\begin{array}{cc}
f(\boldsymbol{k})&G\\
G&-f(\boldsymbol{k})
\end{array}\right)},
\end{eqnarray}
where the function $f(\boldsymbol{k})=2J(\cos k_x+\cos k_y)$ is the dispersion of a monolayer square lattice.
After diagonalization and in terms of eigenoperators $\{\hat{u}_{\boldsymbol{k}},\hat{l}_{\boldsymbol{k}}\}$,
$\hat{H}_B=\sum_{\boldsymbol{k}}[\omega_u(\boldsymbol{k})\hat{u}_{\boldsymbol{k}}^\dag\hat{u}_{\boldsymbol{k}}
+\omega_l(\boldsymbol{k})\hat{l}_{\boldsymbol{k}}^\dag\hat{l}_{\boldsymbol{k}}]$,
with the dispersion of hybrid energy bands
\begin{eqnarray}\label{ME3}
\omega_{u,l}(\boldsymbol{k})=\pm\sqrt{f^2(\boldsymbol{k})+G^2}.
\end{eqnarray}
The relation between the $\{\hat{a}_{1\boldsymbol{k}},\hat{a}_{2\boldsymbol{k}}\}$
and polariton operators $\{\hat{u}_{\boldsymbol{k}},\hat{l}_{\boldsymbol{k}}\}$ is given by
\begin{eqnarray}\label{ME4}
{\left( \begin{array}{cc}
\hat{u}_{\boldsymbol{k}}\\
\hat{l}_{\boldsymbol{k}}
\end{array}\right )}=
{\left( \begin{array}{cc}
-\sin\theta_{\boldsymbol{k}} & \cos\theta_{\boldsymbol{k}}\\
\cos\theta_{\boldsymbol{k}} & \sin\theta_{\boldsymbol{k}}
\end{array}\right )}
{\left( \begin{array}{cc}
\hat{a}_{1\boldsymbol{k}}\\
\hat{a}_{2\boldsymbol{k}}
\end{array}\right )},
\end{eqnarray}
with $\sin\theta_{\boldsymbol{k}}=G/\sqrt{G^2+[\omega_{l}(\boldsymbol{k})+f(\boldsymbol{k})]^2}$
and $\cos\theta_{\boldsymbol{k}}=G/\sqrt{G^2+[\omega_{u}(\boldsymbol{k})+f(\boldsymbol{k})]^2}$.
Because of the chiral symmetry protection, the bands are obviously symmetric about zero energy
[$\hat\sigma_y\hat{H}(\boldsymbol{k})\hat\sigma_y=-\hat{H}(\boldsymbol{k})$]
and a middle bandgap is opened with size $\varepsilon_\text{gap}=2G$.
When $\eta\neq-1$, the bath is still bipartite (if looking at a suitably enlarged unit cell),
and the eigenvalues are in pairs $\omega_u(\boldsymbol{k})=-\omega_l(\boldsymbol{k}+\boldsymbol{\Pi})$, with $\boldsymbol{\Pi}=(\pi,\pi)$.
The origin of this gap is the resonance anti-crossing
between the bare energy bands ($G$=0) at the Van-Hove singularities.
The dispersion of hybridized energy bands are shown in Fig.~1(b).
In particular, the shape of the middle band-edges ($\pm G$) is squarelike, where $f(\boldsymbol{k})=0$.

In this work, we focus on the QE-photon interactions within the middle bandgap.
We consider one (or several) two-level systems $\{|e\rangle,|g\rangle\}$
as the QEs which are locally coupled to a single polarization of light of
the bath at one or more sites, with transition frequency $\omega_e$.
Other polarized light is decoupled from the systems.
The free Hamiltonian of these QEs reads $\hat{H}_S=\Delta/2\sum_{j}\hat{\sigma}_z^j$,
with detuning $\Delta=\omega_e-\omega_1$ and spin operators $\{\hat{\sigma}_z,\hat{\sigma}^\dag,\hat{\sigma}\}$.
The QE-bath interaction of a single giant atom (the $j$th spin) can be described as
\begin{eqnarray}\label{ME5}
\hat{H}_\text{int}^j=\hat{\sigma}_j^\dag\Big(\sum_{\alpha=1}^{N_p}g_{\boldsymbol{n}_\alpha}\hat{a}_{1\boldsymbol{n}_\alpha}
+\sum_{\beta=1}^{N_q}g_{\boldsymbol{n}_\beta}\hat{a}_{2\boldsymbol{n}_\beta}\Big)+\text{H.c.},
\end{eqnarray}
where $g_{\boldsymbol{n}_{\alpha,\beta}}$ is the QE-bath coupling strength at certain coupling points,
and $N_p$ ($N_q$) is the number of coupling points in the top (bottom) layer.
When choosing $\{N_p,N_q\}=\{1,0\}$ or $\{0,1\}$, the interaction recovers the result of a small atom.
This QE-bath interaction Hamiltonian can be transformed into $\boldsymbol{k}$ space
\begin{eqnarray}\label{ME6}
\hat{H}_\text{int}^j&=&\frac{1}{\sqrt{N}}\hat{\sigma}_j^\dag\sum_{\boldsymbol{k}}
\Big[\sum_{\alpha=1}^{N_p}g_{\boldsymbol{n}_\alpha}e^{i\boldsymbol{k}\cdot\boldsymbol{n}_\alpha}
(-\sin\theta_{\boldsymbol{k}}\hat{u}_{\boldsymbol{k}}+\cos\theta_{\boldsymbol{k}}\hat{l}_{\boldsymbol{k}})\notag\\
&+&\sum_{\beta=1}^{N_q}g_{\boldsymbol{n}_\beta}e^{i\boldsymbol{k}\cdot\boldsymbol{n}_\beta}
(\cos\theta_{\boldsymbol{k}}\hat{u}_{\boldsymbol{k}}+\sin\theta_{\boldsymbol{k}}\hat{l}_{\boldsymbol{k}})\Big]+\text{H.c.}.
\end{eqnarray}
Thus, the Hamiltonian of the whole system can be written as
$\hat{H}=\hat{H}_B+\hat{H}_S+\sum_j\hat{H}_\text{int}^j$.

\section{Implementations}

Before proceeding, we make a simple discussion about possible physical implementations
for examining the single-particle physics of our Hamiltonian model, in which particle statistics plays a negligible role.
As we ignore the polarization degree of freedom,
the double-layer square lattices (i.e. the bath Hamiltonian) can be realized in various platforms,
such as cold atoms in optical lattices \cite{Gall2021Competing,Meng2023Atomic}
and optomechanical systems \cite{PhysRevLett.111.073603,PhysRevX.5.031011}.
The two-level emitters (small atoms) can be deeply trapped atom internal states
\cite{PhysRevLett.101.260404,Navarrete-Benlloch_2011,PhysRevA.109.023306} and color centers
\cite{Controlling2018Sohn,PhysRevLett.120.213603,PhysRevB.97.205444,PhysRevX.8.041027}, respectively.
Taking cold atom realizations as an example,
we may consider alkaline-earth-metal atoms with different internal states trapped by optical potentials
\cite{PhysRevLett.101.170504},
which mimic the bilayer square lattices and the QE behaviour.
The bath Hamiltonian can be obtained and controlled using laser-assisted tunneling
\cite{PhysRevLett.107.255301,PhysRevLett.111.185301,PhysRevLett.111.185302,PhysRevLett.114.225301,PhysRevLett.131.080401},
while an empty bath corresponds to the Fermi level $E_F=\text{min}[\omega_{\boldsymbol{k}}]$.
For typical temperatures $T$ of a few $100$ nK and hopping rates $J/2\pi\sim$ kHz
\cite{Sonderhouse2020Thermodynamics,PhysRevLett.111.185301},
the initial mean thermal excitation number obeying Fermi-Dirac distribution is about
$\langle\hat{a}_{\boldsymbol{k}}^\dag\hat{a}_{\boldsymbol{k}}\rangle
=(1+e^{\hbar(\omega_{\boldsymbol{k}}-E_F)/k_BT})^{-1}\sim0.1$ for modes near the Van-Hove singularities.
Lower temperature fulfilling $k_BT\ll\hbar G$ is preferable.
Moreover, matter-wave waveguide QED setups have already been reported in experiments
\cite{Krinner2018Spontaneous,PhysRevResearch.2.043307}.
Therefore, we can restrict ourselves to the ideal case of $T=0$ in the following sections.
As for the impurities, a theoretical proposal for engineering giant atoms
in 2D dynamical state-dependent optical lattices has recently been put forward
\cite{PhysRevLett.122.203603}.

\section{Bound states with small atoms}

\begin{figure}[t]
\includegraphics[scale=0.2]{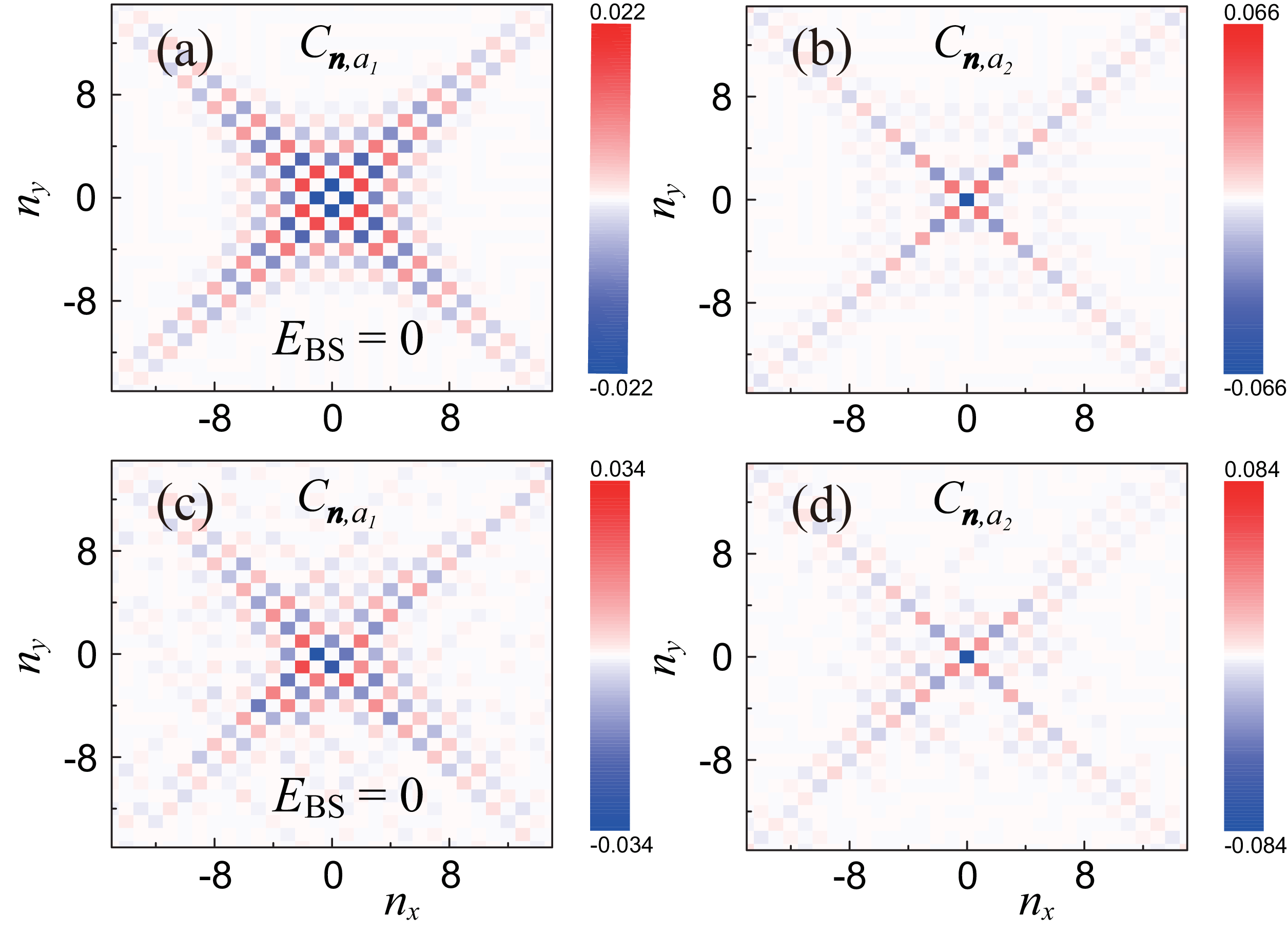}
\caption{\label{fig2}
Spatial distributions (a,c) $C_{\boldsymbol{n},a_1}$ and (b,d) $C_{\boldsymbol{n},a_2}$
of the odd-neighbor, robust and anisotropic BSs with $G=J/4$,
when the emitter with interaction strength $g=0.1J$ is coupled to the $\boldsymbol{n}=0$ site in the top layer.
There are no disorders in (a,b) and off-diagonal disorders in (c,d).}
\end{figure}

In this work, we are particularly interested in the BS produced in the inner bandgap of manufactured bilayer structures.
We first consider a small atom coupled to the site $\boldsymbol{n}_0=(0,0)$ in the top layer, with the Hamiltonian
$\hat{H}_\text{sys}\!=\!\Delta/2\hat{\sigma}_z+g(\hat{a}_{1\boldsymbol{n}_0}\hat{\sigma}^\dag+\text{H.c.})$.
Then we can calculate the eigenvalue and the wave function by solving the stationary Schr{\"o}dinger equation
$(\hat{H}_B+\hat{H}_\text{sys})|\psi\rangle=E_\text{BS}|\psi\rangle$ in momentum space,
with a general one-particle state
\begin{eqnarray}\label{ME7}
|\psi\rangle=\Big(C_e\hat{\sigma}^\dag+1/\sqrt{N}\sum_{\boldsymbol{k}}\sum_{\beta=a_1,a_2}\!\!\!
C_{\boldsymbol{k},\beta}\beta_{\boldsymbol{k}}^\dag\Big)|g\rangle|0\rangle.
\end{eqnarray}
Here $|0\rangle$ represents the ground state of the bath,
and the coefficients $C_e$, $C_{\boldsymbol{k},a_1}=-gC_ef(\boldsymbol{k})/\omega^2(\boldsymbol{k})$
and $C_{\boldsymbol{k},a_2}=[G/f(\boldsymbol{k})]C_{\boldsymbol{k},a_1}$ are the probability amplitudes.
Though the energy $E_\text{BS}\!\in\!(-G,G)$ can be exactly solved by the pole equation: $E_\text{BS}\!=\!\Delta+\Sigma_e(E_\text{BS})$, with self-energy $\Sigma_e$
\cite{doi:10.1002/sca.4950140612},
$E_\text{BS}\!\approx\!\Delta$ is a good approximation when $\{g,\Delta\}\ll G$.
In particular, $E_\text{BS}\!=\!\Delta$ when the spin is resonant with the cavity mode.
Finally, we obtain the spatial distribution of the photonic part of the BS by means of the inverse Fourier transform $C_{\boldsymbol{n},a_{1(2)}}\propto\sum_{\boldsymbol{n}}e^{i\boldsymbol{k}\cdot\boldsymbol{n}}C_{\boldsymbol{k},a_{1(2)}}$.

Different from the common BSs induced by the band-edge at
$\boldsymbol{k}_\text{edge}=(0,0)$ or $\pm(\pi,\pm\pi)$ in a monolayer square lattice
\cite{Subwavelength2015AGT},
the BS shown in parallel arrays is derived from the interactions with the band-edges
around $\boldsymbol{k}_\text{edge}=\pm(k_x,\pi\pm k_x)$ [see Fig.~1(b)].
Returning to real space, the superposition of exponentials resulting from interactions with the opposite side of the square band-edges can provide an interference effect, which causes the wave function to be enhanced or suppressed in certain neighboring sites.
The maximum interference is observed when the QE's transition frequency is at the middle of the bandgap.

In Fig.~2, we plot the spatial distribution of the photonic part of the BSs,
when a single small atom with frequency $\Delta=0$ is coupled to the $\boldsymbol{n}=0$ site in the top layer.
We list the main features of these BSs as follow:
First, in the top layer, the wave function only emerges at the odd lattices, while in the bottom layer, it appears at the even lattices.
We denote this exotic dressed state odd-neighbor BS because they are both odd-neighbor sites with regard to the site to which the QE is coupled. Second, the wave function phases alternate with $\pm1$, with the exception of the vicinity around the QE. Third, the BS has an anisotropic nature
\cite{Subwavelength2015AGT,PhysRevLett.122.203603},
with the wave function predominantly distributed along the $n_x\pm n_y=\pm(\mp)1$ directions in the top layer and along the $n_x\pm n_y=0$ directions in the bottom layer, due to the square-like dispersion of the band-edges.

Another interesting feature of these odd-neighbor BSs is
the robustness to off-diagonal disorders (i.e., disorders in the hopping amplitudes).
To illustrate this effect, we plot odd-neighbor BSs under the off-diagonal disorders in Figs.~2(c,d),
where we add random perturbations $\hat{H}_\text{dis}=\sum_{<\boldsymbol{n},\boldsymbol{m}>}(
\varepsilon_{1\boldsymbol{n}}\hat{a}_{1\boldsymbol{n}}^\dag\hat{a}_{1\boldsymbol{m}}
+\varepsilon_{2\boldsymbol{n}}\hat{a}_{2\boldsymbol{n}}^\dag\hat{a}_{2\boldsymbol{m}}
+\varepsilon_{3\boldsymbol{n}}\hat{a}_{1\boldsymbol{n}}^\dag\hat{a}_{2\boldsymbol{n}}+\text{H.c.})$
to the bath Hamiltonian $\hat{H}_B$, with disorder strengths
$\{\varepsilon_{1\boldsymbol{n}},\varepsilon_{2\boldsymbol{n}}\}\in[-J/4,J/4]$
and $\varepsilon_{3\boldsymbol{n}}\in[-G/4,G/4]$.
With respect to the $\boldsymbol{n}=0$ site, we discover that there is no distribution in the even-neighbor sites, and the BSs' energy is fixed at zero.

To further understand the odd-neighbor profiles and the robustness,
we turn to consider a unit cell including more sites such that the bath Hamiltonian is bipartite.
When the QEs with transition frequency $\Delta=0$ only coupling to the odd sites defined by
$\text{sum}(\boldsymbol{n})\equiv(n_x+n_y)\in\mathbb{Z}_\text{odd}$,
the full Hamiltonian in terms of the block matrix can be formally written as
\begin{eqnarray}\label{ME8}
\bm{H}={\left(\begin{array}{ccc}
0&\bm{Q}_p&\bm{g}\\
\bm{Q}_p^\dag &0&0\\
\bm{g}^\dag &0&0
\end{array}\right)},
\end{eqnarray}
with the submatrix $\bm{Q}_p$ describing the bipartite interactions and
the submatrix $\bm{g}$ being the atom-photon coupling term in odd-site subspace.
By solving $\bm{H}\bm\Psi=E\bm\Psi$ with eigenvector $\bm\Psi=(\bm\Psi_{odd}^T, \bm\Psi_{even}^T, \bm\Psi_{QE}^T)^T$,
we can find an eigenvalue $E=0$ with $\bm\Psi_{odd}=0$ for this block matrix since $\text{det}(\boldsymbol{H})=0$.
Obviously, $E=0$ persists when only adding off-diagonal disorders $d\bm{Q}_p$ to the lattice,
while $\bm\Psi_{odd}=0$ manifests vanishing wave function in odd-site subspace (i.e., odd-neighbor property).
This general conclusion can be applied to arbitrary dimensions.
Indeed, this phenomenon has already been observed in the 1D and 3D chiral symmetric photonic lattices
\cite{Belloeaaw0297,PhysRevLett.126.203601,PhysRevLett.125.163602}.

\section{Bound states with giant atoms}

\begin{figure}[t]
\includegraphics[scale=0.2]{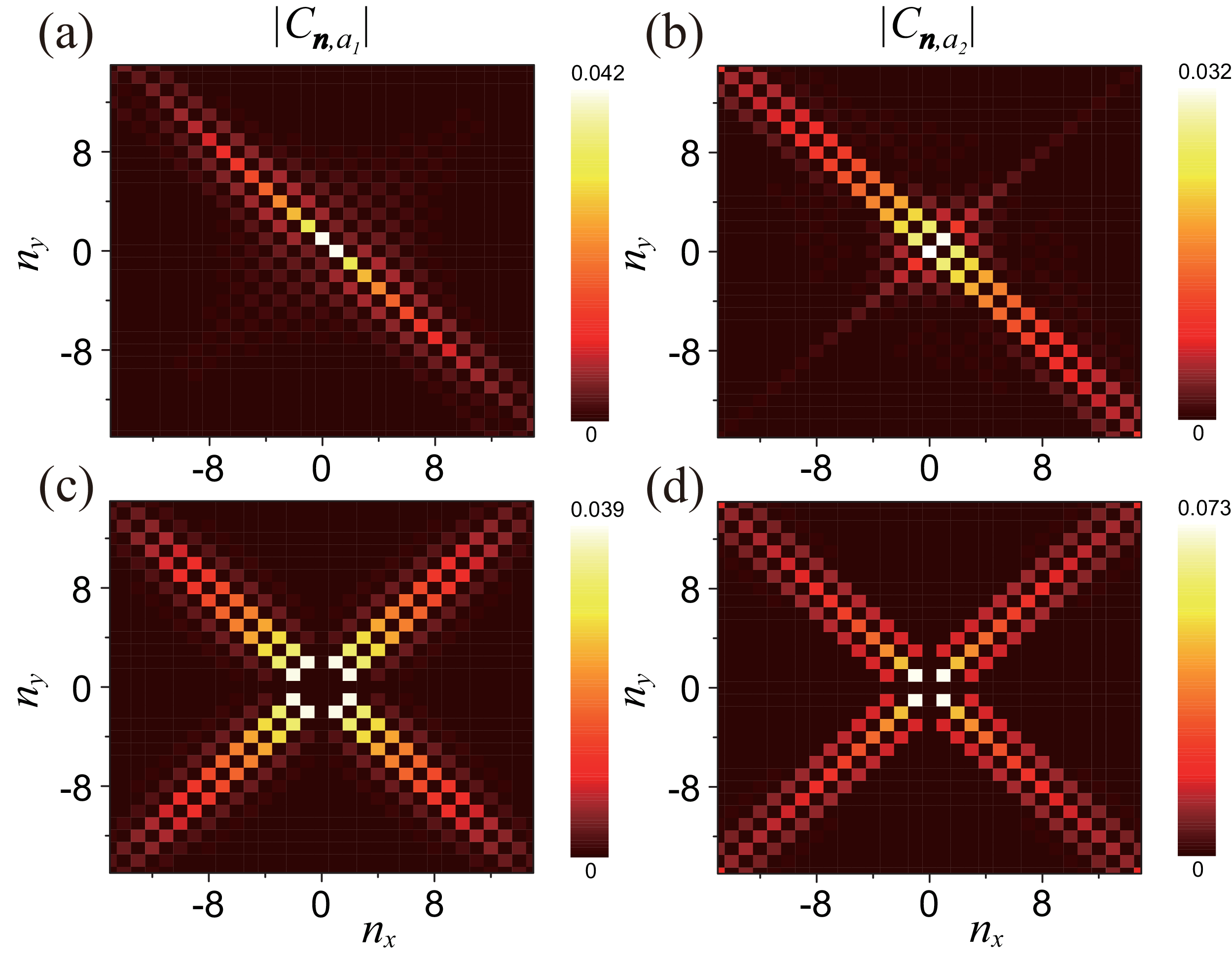}
\caption{\label{fig3}
Wave function distribution of the odd-neighbor BSs formed by a giant atom coupling to bilayer square arrays.
(a,b) The giant atom is coupled to two sites $\boldsymbol{n}_{1,2}=(0,0)$ and $(1,1)$ in layer 1.
(c,d) The giant atom is coupled to four sites $\boldsymbol{n}=(\pm1,\pm1),(\pm1,\mp1)$ in layer 1.}
\end{figure}

The pattern of the BSs can be further engineered by taking advantage of giant atoms.
To protect the BSs' odd-neighbor property, the coupling sites of a single giant atom should be even-neighbor.
Quantum interference between nonlocal coupling points can result in the cancellation of
the spatial distribution of the BSs in certain lattice sites or directions.

We first consider the case when a giant atom is coupled to
two sites $\boldsymbol{n}_{1,2}=(0,0)$ and $(1,1)$ in the top layer with identical coupling strength $g$.
Then the QE-bath interaction Hamiltonian in $\boldsymbol{k}$ space simplifies to
\begin{eqnarray}\label{ME9}
\hat{H}_\text{int}&=&\frac{g}{\sqrt{N}}\hat{\sigma}^\dag\sum_{\boldsymbol{k}}[1+e^{i(k_x+k_y)}]\notag\\
&\times&(-\sin\theta_{\boldsymbol{k}}\hat{u}_{\boldsymbol{k}}+\cos\theta_{\boldsymbol{k}}\hat{l}_{\boldsymbol{k}})+\text{H.c.},
\end{eqnarray}
where the interference term $\text{I}(\boldsymbol{k})=1+e^{i(k_x+k_y)}$ results in destructive interferences
for band-edges modes $\boldsymbol{k}_\text{edge}=\pm(k_x,\pi-k_x)$.
This is reflected in the cancellation of one branch of the anisotropic BS pattern, as shown in Figs.~3(a,b).
Another example is to homogenize the anisotropy of the BS through coupling to four positions in the top layer.
The four sites are $\boldsymbol{n}=(\pm1,\pm1),(\pm1,\mp1)$ and the interference term
in $\boldsymbol{k}$ space can be chosen as $\text{I}(\boldsymbol{k})=4\sin k_x\sin k_y$.
Clearly, for the band-edge modes $\boldsymbol{k}_\text{edge}=(0,\pm\pi)$ and $(\pm\pi,0)$, this term approaches zero.
As a result, the non-Markovian part in the BS stemmed from Van-Hove singularities is filtered,
as plotted in Figs.~3(c,d).

\section{Bipartite spin model}

\begin{figure}[t]
\includegraphics[scale=0.19]{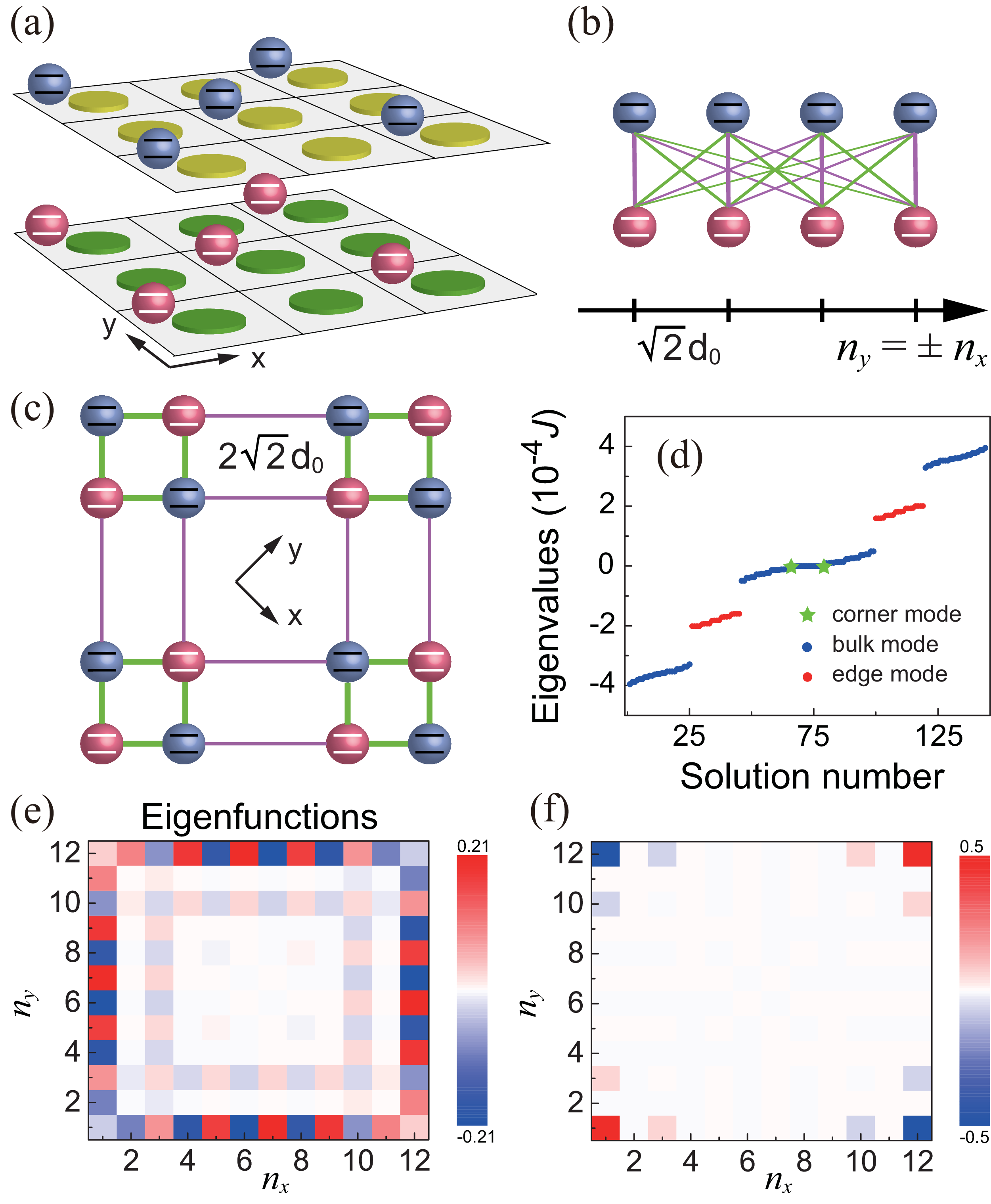}
\caption{\label{fig4}
(a) Bipartite spin model with all the QEs at even lattices,
that is $\text{sum}(\boldsymbol{n})\in\mathbb{Z}_\text{even}$.
The QEs in layer 1 (blue) are only coupled to
the QEs in layer 2 (red) and vice versa.
(b) Spin interactions of (a) along the diagonal directions.
(c) Vertical view of the configuration of a generalized 2D SSH model in the trivial phase,
with QEs alternately and dimerically placed in the double layers.
(d) Spin model energy spectrum of a finite system with 144 small atoms inside the $35\times35$ bilayer square arrays.
(e,f) One of the edge modes and corner modes extracted from the spectrum of
the second-order topological insulator shown in (d). Here, we choose $G=4J$ and $g=0.1J$.}
\end{figure}

The odd-neighbor, robust and anisotropic BSs can mediate long-range tunable, chiral symmetric
and anisotropic spin-spin interactions when involving multiple QEs.
In the Markovian limit, the coherent interactions can be described as
$\hat{H}_S=\sum_{i<j}(g_{ij}\hat{\sigma}_i^\dag\hat{\sigma}_j+\text{H.c.})$,
where the coupling strength $g_{ij}$ vanishes for $\text{sum}(\boldsymbol{n}_{ij})\in\mathbb{Z}_\text{even}$
($\mathbb{Z}_\text{odd}$) when two QEs are in the same (different) layer(s),
with relative position $\boldsymbol{n}_{ij}=\boldsymbol{n}_j-\boldsymbol{n}_i$.
In fact, a general analysis in Ref.~\cite{ro2024hermitian} has shown that the emitter Hamiltonian can inherit chiral symmetry of the photonic bath
in one-emitter-per-resonator case, when setting $\Delta=0$.
A direct application of this spin model can realize high-fidelity entanglement of multiple QEs located at the odd lattices
via an auxiliary one at the even lattice, where the unwanted crosstalk is reduced
\cite{doi:10.1021/acsphotonics.8b01455}.

A potentially interesting configuration based on parity properties is
that the photonic lattice is half-filled by the QEs,
in which the QEs only interact if they are coupled to the lattice sites in different layers,
as shown schematically in Figs.~4(a,b).
The anisotropic spin-spin interactions are mainly along the diagonal directions of the lattice arrays [see Fig.~4(b)].
At this time, the spin model is bipartite in the layer space, with the Bloch Hamiltonian
\begin{eqnarray}\label{ME10}
\hat{H}_S(\boldsymbol{k})={\left(\begin{array}{cc}
0&f_S(\boldsymbol{k})\\
f_S^*(\boldsymbol{k})&0
\end{array}\right)},
\end{eqnarray}
where $f_S(\boldsymbol{k})=\sum_{\boldsymbol{n}}
J_{1,2}^{\boldsymbol{n}}\exp(-i\boldsymbol{k}\cdot\boldsymbol{n})$ is the off-diagonal term
and $J_{1,2}^{\boldsymbol{n}}$ denotes the coupling between the QEs coming from
the different layers at a distance $\boldsymbol{n}$.
The system also has inversion symmetry $\hat{H}_S(\boldsymbol{k})\rightarrow\hat{H}_S(-\boldsymbol{k})$,
with $\hat\sigma_x$ the corresponding local operator
\cite{PhysRevLett.127.147401}.
The spin model protected by chiral symmetry can be harnessed to simulate exotic many-body phases,
such as double N\'{e}el ordered states
\cite{Belloeaaw0297,PRXQuantum.3.010336}.

\section{Generalized 2D SSH model}
Apart from the key symmetries that provide topological protection,
bond order is another crucial ingredient required for entering topology.
We selectively position the QEs in a dimeric and alternating manner
in the bipartite spin model whose vertical view is plotted in Fig.~4(c).
This specific spin array arrangement can investigate the topological phases of a generalized 2D SSH model
\cite{PhysRevB.102.134213,Leefmans2022Topological,Parto2023,PhysRevLett.118.076803,PhysRevLett.122.233903,
PhysRevB.100.075120,PhysRevB.100.075437,PhysRevApplied.12.034014,Li:22}.

We consider a system composed of
$35\times35$ bilayer square lattices and a $12\times12$ spin array,
with a configuration similar to that shown in Fig.~4(c).
The energy spectrum of the spin interactions is plotted in Fig.~4(d), with $G=4J$.
Here, the inner band gap of $8J$ occupies most of the entire range of $8\sqrt{2}J$.
We point out that the ratio between out-of-plane coupling and in-of-plane coupling $G/J$
mainly influences the localization length of the BSs.
For a given light-matter coupling $g\ll\{G,J\}$, the increasing of the detuning to band-edges
reduces the localization length of the BSs and thereby the spin couplings.
The anisotropy is weakened as well even though $f(\boldsymbol{k})=0$ for $\omega(\boldsymbol{k})=\pm G$ is still satisfied,
as compared to the patterns under weak interlayer coupling $G=J/4$.
However, the odd-neighbor and robust properties stemming from chiral symmetry are independent of the ratio $G/J$.
The profiles of the BSs are more localized such that
we can approximate emitter interactions to the third-neighbor or even nearest-neighbor hoppings.
In this regime, the edge modes are spectrally isolated from the bulk energy bands, as shown in Fig.~4(d).
Despite the challenges, we believe that $10^{-4}J$ for the energy eigenvalues is likely
at finite temperatures in cold atom realizations.
This is because the spin frequencies are far away from the Fermi level of an empty bath,
and the bath modes are traced out for arriving at spin-spin interactions.

By expanding the spin couplings up to the third-neighbor hoppings,
the non-diagonal term $f_S(\boldsymbol{k})$ in Eq.~(\ref{ME10}) can be rewritten as
\begin{eqnarray}\label{ME11}
f_\text{SSH}(\bar{\boldsymbol{k}})\simeq
{\left(\begin{array}{cc}
f_0(\bar{k}_x)&f_0(\bar{k}_y)\\
f_0^*(\bar{k}_y)&f_0^*(\bar{k}_x)
\end{array}\right)},
\end{eqnarray}
with $f_0(\bar{k}_j)=t_1+t_2e^{i\bar{k}_j}+t_3e^{-i\bar{k}_j}+t_4e^{2i\bar{k}_j}$ ($j=x,y$)
and wave vector $\bar{\boldsymbol{k}}=(\bar{k}_x,\bar{k}_y)$ for spin array.
In the topological phase, one finds $|t_2|\gg|t_1|\gg|t_4|\gg|t_3|$.
To characterize the existence of edge modes, we should calculate the topological polarization through the integral
\cite{PhysRevLett.118.076803}
\begin{eqnarray}\label{ME12}
\mathbf{P}=\frac{1}{2\pi}\int_\text{BZ}d\bar{\boldsymbol{k}}\text{Tr}[\mathcal{A}_m(\bar{\boldsymbol{k}})],
\end{eqnarray}
where $\mathcal{A}_m(\bar{\boldsymbol{k}})=i\psi_m^\dag(\bar{\boldsymbol{k}}) \partial_{\bar{\boldsymbol{k}}}\psi_m(\bar{\boldsymbol{k}})$
is the non-Abelian Berry connection and $\psi_m(\bar{\boldsymbol{k}})$
represents the eigenfunction of spin Hamiltonian with band index $m$.
In addition to numerical calculation, another method is based on symmetry analysis
\cite{PhysRevB.99.245151}.
Because the spin model respects $C_4$ symmetry,
the values of the polarization are either $(0,0)$ or $(\frac{1}{2},\frac{1}{2})$
for system being in the trivial or topological phase, respectively.
The existence of corner states can be further characterized by the secondary topological index $Q_\text{corner}=\frac{1}{4}$.
Note that the corner modes here are also known as bound states in the continuum
with the protection of the simultaneous $C_{4v}$ and sublattice symmetries.
A detailed analysis can be found in Ref.~\cite{PhysRevB.101.161116}.
In Fig.~4(e), we extract and plot one of the edge states distributed along the boundary of the 2D spin array.
While in Fig.~4(f), we give a topological corner state with distributions localized at the four corners of the square array,
a direct evidence of nontrivial second-order topological insulating phases
\cite{PhysRevB.98.245413,PhysRevLett.122.076801}.

\section{Conclusion}
In summary, we developed a unique photonic bath with a symmetric middle bandgap caused by parallel array interlayer hybridization.
In this bandgap, the quantum optical properties of small and giant atoms are investigated.
Interactions with squarelike band-edge modes of a chiral symmetric photonic lattice, in particular, result in odd-neighbor, robust, and anisotropic BSs, which can mediate novel spin-spin interactions that inherit the BSs' properties.
We have demonstrated the capability of our system in the simulation of nontrivial many-body phases, such as
topological phases of high-order topological insulators.

\begin{acknowledgments}
We would like to thank Dr. Zongping Gong for fruitful discussions.
The simulations are obtained using QuTiP
\cite{JOHANSSON20121760,JOHANSSON20131234}.
P.B.L. is supported by the National Natural Science
Foundation of China under Grant No. 92065105, and the Natural Science Basic Research Program
of Shaanxi (Program No. 2020JC-02).
F.N. is supported in part by Nippon
Telegraph and Telephone Corporation (NTT) Research,
Japan Science, and Technology Agency (JST) (via the
Quantum Leap Flagship Program (Q-LEAP), Moonshot
R\&D Grant No. JPMJMS2061, Army Research Office (ARO)
(Grant No. W911NF-18-1-0358), the Asian Office of
Aerospace Research and Development (AOARD) (via
Grant No. FA2386-20-1-4069), and the Foundational
Questions Institute (FQXi) (via Grant No. FQXiIAF19-06).
\end{acknowledgments}

\appendix
\section{Bilayer square lattices of $\eta\neq-1$}

\begin{figure}[t]
\includegraphics[scale=0.19]{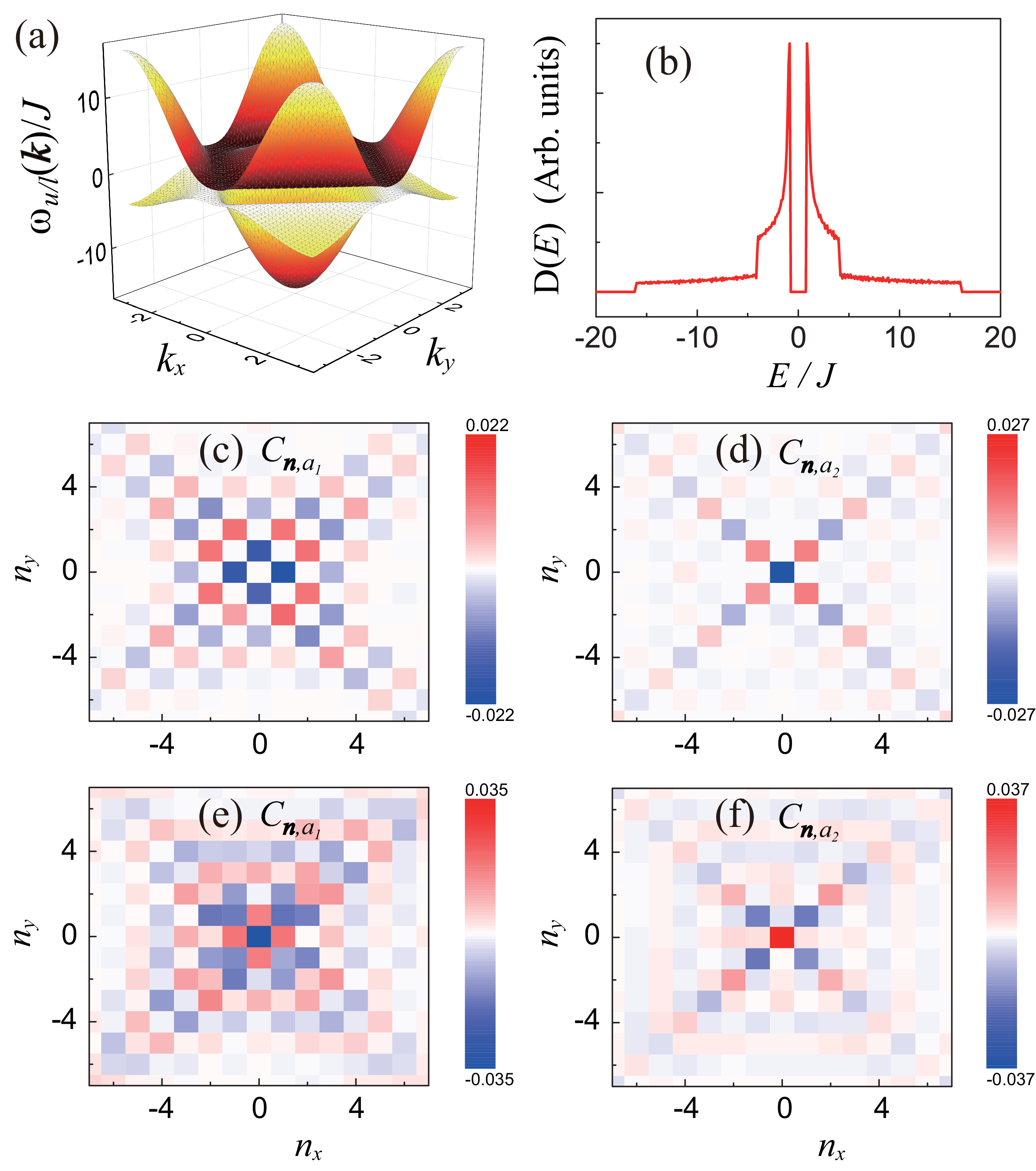}
\caption{\label{fig5}
(a) Dispersion relationship of hybrid energy bands.
(b) Corresponding density of states.
(c-f) Wavefunction spatial distribution of the BS with a small random off-diagonal disorder.
In (c,d) $\Delta=0$ and (e,f) $\Delta=0.5J$.
Here, $\eta=-4$, $G=J$ and $g=0.1J$.}
\end{figure}

Although we particularly focus on the $\eta=-1$ case in the main text,
we present some results in the $\eta\neq-1$ case in this appendix.
The dispersion relations of two hybrid energy bands read
\begin{eqnarray}
\omega_{u/l}(\boldsymbol{k})=\frac{(1+\eta)f(\boldsymbol{k})}{2}\pm\sqrt{\frac{(1-\eta)^2f^2(\boldsymbol{k})}{4}+G^2}.
\end{eqnarray}
It is obvious that the solutions are no longer strictly symmetric
$\boldsymbol{k}\rightarrow-\boldsymbol{k}$ with respect to the zero energy,
but still come in pairs $\omega_u(\boldsymbol{k'})=-\omega_l(\boldsymbol{k})$, with $\boldsymbol{k'}=(k_x+\pi,k_y+\pi)$.
Besides, the energy gap opened around Van-Hove singularities persists as long as $G\neq0$ and $\eta<0$,
even though the band-edge modes no longer entirely satisfy $f(\boldsymbol{k})=0$.
Nevertheless, the shape of the middle band-edges is still square-like.
To clarify it, we plot the energy bands in Fig.~5(a), with $\eta=-4$ and $G=J$.
The corresponding density of states is also given in Fig.~5(b), which shows a divergency in the middle band-edges.

\section{Details of bound states with a single coupling point}
As discussed in the main text, the full Hamiltonian is given by $\hat{H}=\hat{H}_B+\hat{H}_\text{sys}$.
The QE-photon dressed state formed in the single-excitation subspace can be solved based on the secular equation
$\hat{H}|\psi\rangle=E_{BS}|\psi\rangle$, with the eigenvalue $E_\text{BS}$ and one particle ansatz
\begin{eqnarray}
|\psi\rangle=\Big(C_e\hat{\sigma}^\dag+\sum_{\boldsymbol{n}}\sum_{\beta=a_1,a_2}
C_{\boldsymbol{n},\beta}\beta_{\boldsymbol{n}}^\dag\Big)|g\rangle|0\rangle.
\end{eqnarray}
To obtain the coefficients, we solve the stationary Schr{\"o}dinger equation in $\boldsymbol{k}$ space
and do the inverse Fourier transform, which yields
\begin{eqnarray}
C_{\boldsymbol{n},a_1}&=&\frac{gC_e}{4\pi^2}\iint_\text{BZ}d\boldsymbol{k}e^{i\boldsymbol{k}\cdot\boldsymbol{n}}
C_{\boldsymbol{k},a_1}\\
C_{\boldsymbol{n},a_2}&=&\frac{gC_e}{4\pi^2}\iint_\text{BZ}d\boldsymbol{k}e^{i\boldsymbol{k}\cdot\boldsymbol{n}}
C_{\boldsymbol{k},a_2},
\end{eqnarray}
with wavefunctions in reciprocal space
\begin{eqnarray}
C_{\boldsymbol{k},a_1}&=&\frac{\sin^2\theta_{\boldsymbol{k}}}{E_{\text{BS}}-\omega_u(\boldsymbol{k})}
+\frac{\cos^2\theta_{\boldsymbol{k}}}{E_{\text{BS}}-\omega_l(\boldsymbol{k})}\\
C_{\boldsymbol{k},a_2}&=&\frac{-\sin\theta_{\boldsymbol{k}}\cos\theta_{\boldsymbol{k}}}{E_{\text{BS}}-\omega_u(\boldsymbol{k})}
+\frac{\sin\theta_{\boldsymbol{k}}\cos\theta_{\boldsymbol{k}}}{E_{\text{BS}}-\omega_l(\boldsymbol{k})}
\end{eqnarray}
Here, $\sin\theta_{\boldsymbol{k}}=G/\sqrt{G^2+[\omega_{l}(\boldsymbol{k})-\eta f(\boldsymbol{k})]^2}$
and $\cos\theta_{\boldsymbol{k}}=G/\sqrt{G^2+[\omega_{u}(\boldsymbol{k})-\eta f(\boldsymbol{k})]^2}$.
Though the probability amplitude $C_e$ can be obtained by imposing the normalization condition,
$C_e\approx1$ is a good approximation in the Markovian limit $g\ll\varepsilon_\text{gap}/2$,
with $\varepsilon_\text{gap}$ the size of the middle bandgap.
When considering $\eta=-1$ and $E_\text{BS}=0$, we arrive at
\begin{eqnarray}
C_{\boldsymbol{n},a_1}
&=&\frac{gC_e}{4\pi^2}\iint_\text{BZ}d\boldsymbol{k}e^{i\boldsymbol{k}\cdot\boldsymbol{n}}
\frac{-f(\boldsymbol{k})}{f^2(\boldsymbol{k})+G^2}\\
C_{\boldsymbol{n},a_2}
&=&\frac{gC_e}{4\pi^2}\iint_\text{BZ}d\boldsymbol{k}e^{i\boldsymbol{k}\cdot\boldsymbol{n}}
\frac{-G}{f^2(\boldsymbol{k})+G^2},
\end{eqnarray}
which are used in the main text.
One can easily check that $C_{\boldsymbol{n},a_1}=0$ for $\text{sum}(\boldsymbol{n})\equiv(n_x+n_y)\in\mathbb{Z}_\text{even}$
and $C_{\boldsymbol{n},a_2}=0$ for $\text{sum}(\boldsymbol{n})\in\mathbb{Z}_\text{odd}$
due to the interference between the $\boldsymbol{k}$ mode and the $\boldsymbol{k'}$ mode,
with $\boldsymbol{k'}=(k_x+\pi,k_y+\pi)$.
The intrinsic chiral symmetry of the bath Hamiltonian is responsible for the odd-neighbor profiles and the robustness.

To clarify it, we consider a chiral symmetric system with multiple quantum emitters.
The full Hamiltonian in terms of the block matrix can take the form
\begin{eqnarray}
\bm{H}={\left(\begin{array}{cc}
\bm{H}_p&\bm{g}_{oe}\\
\bm{g}_{oe}^\dag &\bm{\Delta}
\end{array}\right)}, \quad\text{with}\quad
\bm{H}_p={\left(\begin{array}{cc}
0&\bm{Q}_p\\
\bm{Q}_p^\dag &0
\end{array}\right)}
\end{eqnarray}
Here, the submatrix $\bm{H}_p$ is the bath's Hamiltonian, $\bm{g}_{oe}=(\bm{g}_{odd}, \bm{g}_{even})$ is the coupling term
and $\bm\Delta/\Delta=\bm{I}$.
Assuming $\bm\Delta=0$ and no emitters in even sites $\bm{g}_{even}=0$, and using the formula for the block matrices
\begin{eqnarray}
\text{det}{\left(\begin{array}{cc}
\bm{A}&\bm{B}\\
\bm{C}&\bm{D}
\end{array}\right)}=
\text{det}(\bm{A})\text{det}(\bm{D}-\bm{C}\bm{A}^{-1}\bm{B}),
\end{eqnarray}
the determinant of the full Hamiltonian is zero, that is $\text{det}(\bm{H})=0$.
By solving $\bm{H}\bm\Psi=E\bm\Psi$ with eigenvector $\bm\Psi=(\bm\Psi_{odd}^T, \bm\Psi_{even}^T, \bm\Psi_{QE}^T)^T$,
we can always find an eigenvalue $E=0$ with $\bm\Psi_{odd}=0$,
in the presence of off-diagonal disorders $d\bm{Q}_p$.

In Figs.~7(c,d), we plot the photonic part of the BS with $\eta=-4$, $G=J$ and $\Delta=0$.
We still find odd-neighbor and robust properties when off-diagonal disorders are included.
As a contrast, we give the BS pattern in Figs.~7(e,f) with $\Delta=0.5J$,
in which the full Hamiltonian (the emitter part together with the lattice part) is no longer protected by chiral symmetry.
As a result, the single-emitter BS can have weights in both odd and even sites.

\section{Bound states with multiple coupling points}

\begin{figure}[t]
\includegraphics[scale=0.19]{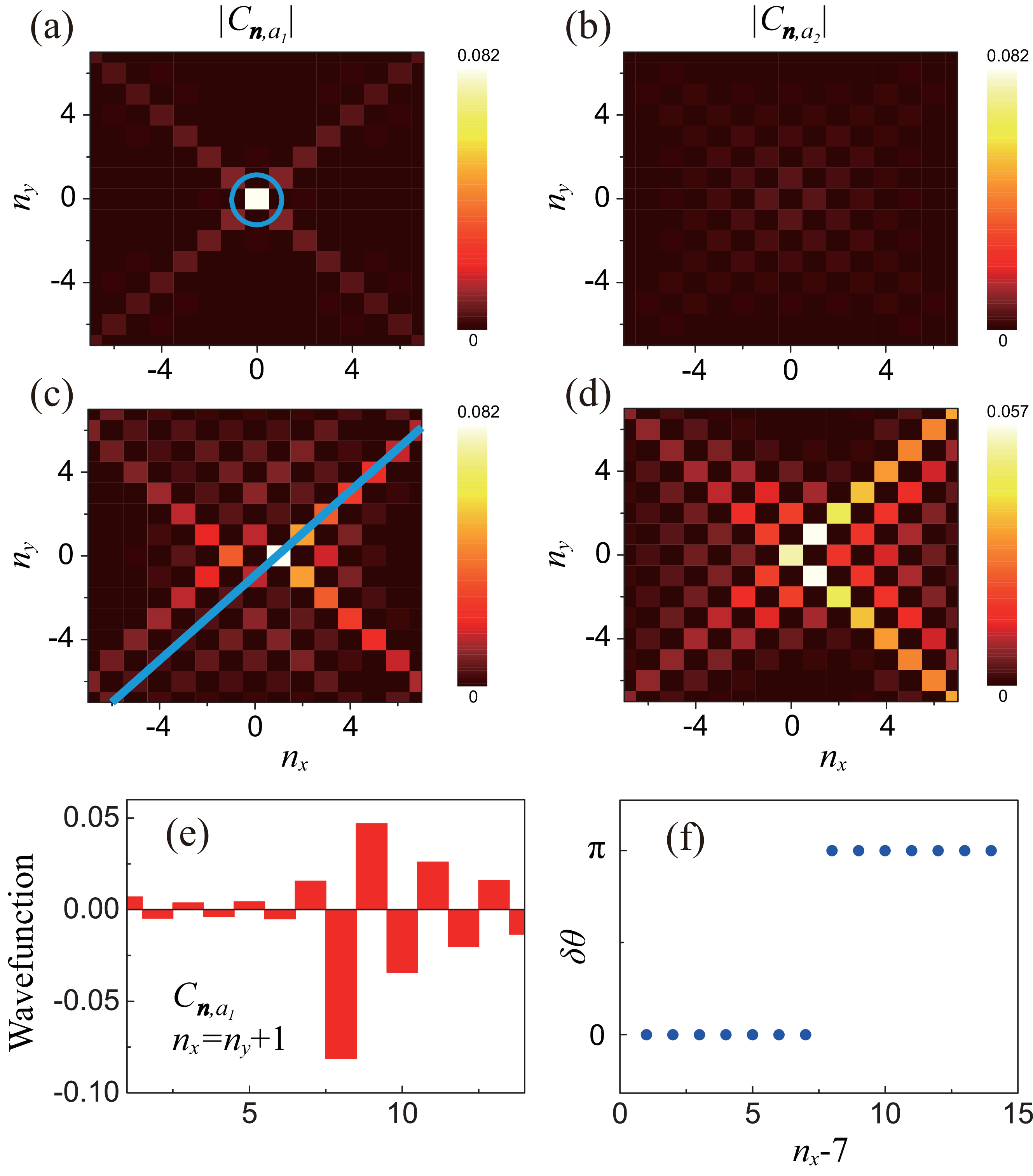}
\caption{\label{fig6}
Modulus of the wavefunction distribution in real space.
(a), (b) Four coupling points $\boldsymbol{n}_{1,2,3,4}=(\pm1,0)/(0,\pm1)$
distributing in the top layer, with identical coupling strength $g$. Photon population is trapped.
(c), (d) Two coupling points $\boldsymbol{n}_{1,2}=(0,0)/(1,0)$ respectively
distributing in the two layers, with identical coupling strength $g$. Chiral patterns occur in some directions.
(e) Spatial distribution $C_{\boldsymbol{n},a_1}$ and (f) phase difference $\delta\theta=\theta_1-\theta_2$
along the $n_x=n_y+1$ direction [labelled by blue line in (c)].
A sudden jump of $\delta\theta$ occurs at $\boldsymbol{n}_2$. Here, $\eta=-1$, $G=J/4$ and $g=0.1J$.}
\end{figure}

\subsection{Some details}

We now detail the results of light-matter interaction beyond dipole approximation.
In this case, the so-called ``giant atom" features nonlocal coupling to the optical lattices.
In terms of intralayer interference factors
\begin{eqnarray}\label{ME16}
\text{I}(\boldsymbol{k})=\sum_{\alpha=1}^{N_p}g_{\boldsymbol{n}_\alpha}e^{i\boldsymbol{k}\cdot\boldsymbol{n}_\alpha},\quad
\text{I}'(\boldsymbol{k})=\sum_{\beta=1}^{N_q}g_{\boldsymbol{n}_\beta}e^{i\boldsymbol{k}\cdot\boldsymbol{n}_\beta},
\end{eqnarray}
which reflect the interference effects stemmed from multiple coupling points of the giant atom,
the interaction Hamiltonian in $\boldsymbol{k}$ space can be rewritten as
\begin{eqnarray}
\hat{H}_\text{int}&=&\frac{1}{\sqrt{N}}\hat{\sigma}^\dag\sum_{\boldsymbol{k}}\big[\text{I}(\boldsymbol{k})
\times(-\sin\theta_{\boldsymbol{k}}\hat{u}_{\boldsymbol{k}}+\cos\theta_{\boldsymbol{k}}\hat{l}_{\boldsymbol{k}})\notag\\
&+&\text{I}'(\boldsymbol{k})\times(\cos\theta_{\boldsymbol{k}}\hat{u}_{\boldsymbol{k}}+\sin\theta_{\boldsymbol{k}}\hat{l}_{\boldsymbol{k}})\big]+\text{H.c.}.
\end{eqnarray}
From the view of the wavefunction distribution of the BS,
the interference pattern is a superposition of multiple BSs based on small atoms
\begin{eqnarray}
C_{\mathbf{n},a_1}^\text{g}
&=&\sum_{\boldsymbol{n}_\alpha}(g_{\boldsymbol{n}_\alpha}/g)C_{\boldsymbol{n}-\boldsymbol{n}_\alpha,a_1}^\text{s}\notag\\
&+&\sum_{\boldsymbol{n}_\beta}(g_{\boldsymbol{n}_\beta}/g)C_{\boldsymbol{n}-\boldsymbol{n}_\beta,a_2}^\text{s}\\
C_{\boldsymbol{n},a_2}^\text{g}
&=&\sum_{\boldsymbol{n}_\alpha}(g_{\boldsymbol{n}_\alpha}/g)C_{\boldsymbol{n}-\boldsymbol{n}_\alpha,a_2}^\text{s}\notag\\
&-&\sum_{\boldsymbol{n}_\beta}(g_{\boldsymbol{n}_\beta}/g)C_{\boldsymbol{n}-\boldsymbol{n}_\beta,a_1}^\text{s}
\end{eqnarray}
To protect the parity property, we assume that the coupling points are even-neighbor with each other.

\subsection{Other patterns}

In addition to the unconventional patterns of two examples shown in Fig.~3, we present some other results in this appendix.
First, we consider four coupling points $\boldsymbol{n}_{1,2,3,4}=(1,0)/(-1,0)/(0,1)/(0,-1)$ in the top layer
with identical coupling strength $g$.
In this case, the interference factor can be expressed as
\begin{eqnarray}
\text{I}(\boldsymbol{k})&=&2g(\cos k_x+\cos k_y),
\end{eqnarray}
which is zero for all the band-edge modes.
This corresponds to the trap of the photon population around giant atom,
which is plotted in Figs.~6(a) and 6(b).

We may further consider a single giant atom with coupling points distributing in two layers.
For simplicity, we assume that the QE is simultaneously coupled to $\boldsymbol{n}_1=(0,0)$ lattice site of the top layer
and $\boldsymbol{n}_2=(1,0)$ lattice site of the bottom layer, with identical coupling strength $g$.
In this case, $\text{I}(\boldsymbol{k})=g$ and $\text{I}'(\boldsymbol{k})=ge^{i(k_x+k_y)}$ in the interaction Hamiltonian.
In Figs.~6(c)-6(e), we plot the BS pattern of such configuration and show chiral feature in some directions,
such as $n_x=n_y+1$ direction. Next, we attempt to understand how it occurs.
Taking the spatial distribution in the top layer for example,
the expression of $C_{\boldsymbol{n},a_1}^\text{g}$ can be rewritten as
\begin{eqnarray}\label{ME21}
C_{\boldsymbol{n},a_1}^\text{g}
&=&C_{\boldsymbol{n}-\boldsymbol{n}_1,a_1}^\text{s}+C_{\boldsymbol{n}-\boldsymbol{n}_2,a_2}^\text{s}\notag\\
&=&Ae^{i\theta_1}+Be^{i\theta_2},
\end{eqnarray}
where $A/B$ denotes the amplitude and $\theta_1/\theta_2$ is the phase.
We recall that the BS for a small atom has alternating phases $\pm1$ in real space.
Thus, the phase difference $\delta\theta=\theta_1-\theta_2$ plays an important role
in the process of two site interferences.
In Fig.~6(f), we plot the relative phase $\delta\theta$ along the $n_x=n_y+1$ direction
and find a sudden jump between $0$ and $\pi$ phase difference.
This means the appearance of destructive (constructive) interference on the left (right) side,
which accounts for the emergence of chirality.

\section{Eight-particle entanglement}

\begin{figure}[t]
\includegraphics[scale=0.17]{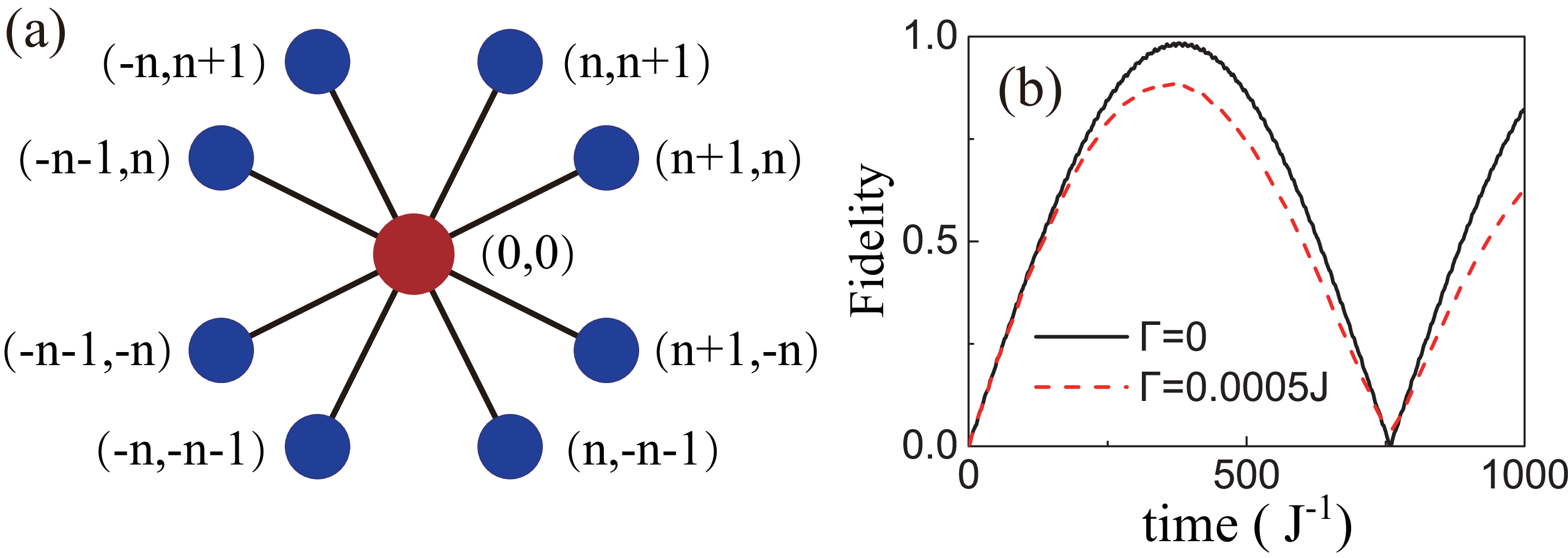}
\caption{\label{fig7}
(a) Configuration for eight-particle entanglement, with nine atoms in the top layer.
Eight of them in the odd sites and an auxiliary one at the origin.
(b) Time evolution of the fidelity of goal state $|\phi\rangle_\text{goal}$,
with initial state $|\phi_0\rangle$.
Both ideal result and the situation under dissipation are plotted.
Here, $\eta=-1$, $G=J/4$ and $g=0.1J$.}
\end{figure}

When a set of small atoms are taken into account,
the BSs can mediate odd-neighbor, robust and anisotropic dipole-dipole interaction.
In the parameter regime of $g\ll G$ (Markovian regime), the effective dynamics is governed by the following Hamiltonian
\begin{eqnarray}
\hat{H}_S=\frac{1}{2}\sum_{i,j}(g_{ij}\hat{\sigma}_i^\dag\hat{\sigma}_j+\text{H.c.})
\end{eqnarray}
with coupling strength
\begin{eqnarray}
g_{ij}=\left\{
\begin{aligned}
&gC_{\boldsymbol{n}_{ij},a_1}/C_e, \!\!\!&Q_i,Q_j\in \text{layer 1}\\
&-gC_{\boldsymbol{n}_{ij},a_1}/C_e, \!\!\!&Q_i,Q_j\in \text{layer 2}\\
&gC_{\boldsymbol{n}_{ij},a_2}/C_e, \!\!\!&Q_i\in \text{layer 1}, Q_j\in \text{layer 2}
\end{aligned}
\right.
\end{eqnarray}
where $\boldsymbol{n}_{ij}=\boldsymbol{n}_{j}-\boldsymbol{n}_{i}$ is the relative position.
Based on parity property of the BS, the coupling strength $g_{ij}$ vanishes for $\text{sum}(\boldsymbol{n}_{ij})\in\mathbb{Z}_\text{even}$
($\mathbb{Z}_\text{odd}$) when two atoms are in the same (different) layers.
This means that we can divide the atoms into two manifolds, and in each manifold there is no intercoupling.

A possible direct application, which makes use of this novel spin-spin interaction,
is long-distance entanglement distribution.
We consider nine QEs located in the top layer of the bilayer lattices,
and eight of them are odd-site QEs along the diagonal directions with
$\boldsymbol{n}_{1-8}=(n,\pm n\pm1)/(-n,\pm n\pm1)/(n+1,\pm n)/(-n-1,\pm n)$,
while an auxiliary one is at the origin, as shown schematically in Fig.~7(a).
The effective Hamiltonian of this setup can be constructed by
\begin{eqnarray}\label{ME24}
\hat{H}_\text{eff}=gC_{\boldsymbol{n}_{1},a_1}/C_e\hat\sigma_a^\dag\sum_{i=1}^{8}\hat\sigma_i+\text{H.c.}
\end{eqnarray}
Here, we assume identical atom-photon coupling strength and in the weak-coupling limit.
In order to prepare the desired entanglement state, we consider the initial state
$|\phi_0\rangle=|e\rangle_a\otimes|g\rangle^{\otimes8}$,
where the auxiliary atom is initially in the excited state
and the other eight atoms are initially in the ground state.
After a time interval $\tau=\pi C_e/(4gC_{\boldsymbol{n}_{1},a_1})$,
the system evolves into the target entanglement state
\begin{eqnarray}\label{ME25}
|\phi\rangle_\text{goal}=|g\rangle_a\otimes\frac{1}{\sqrt{8}}\sum_{i=1}^{8}\hat\sigma_i^\dag|g\rangle^{\otimes8}.
\end{eqnarray}
Compared with previous four-particle entanglement scheme discussed in Ref.~\cite{doi:10.1021/acsphotonics.8b01455},
we can generate an entangled state involving more particles (up to eight),
and more importantly avoid the unwanted cross-talk which dominates the infidelity.
Thus, our scheme may achieve higher fidelity.
The time evolution of the dynamics under dissipation is governed by the following master equation
\begin{eqnarray}
\frac{d\hat\rho}{dt}&=&-i[\hat{H}_{\text{eff}},\hat{\rho}]
+n_\text{ph}\kappa\sum_\mathbf{n}(D[\hat{a}_{1,\boldsymbol{n}}]\hat{\rho}+D[\hat{a}_{2,\boldsymbol{n}}]\hat{\rho})\notag\\
&+&\Gamma\sum_{i=1}^{8}D[\hat{\sigma}_{i}]\hat{\rho}+\Gamma D[\hat{\sigma}_{a}]\hat{\rho},
\end{eqnarray}
where $n_\text{ph}$ is the photonic component of the BS,
$\kappa$ and $\Gamma$ denote the photon loss of each site and the decay rate of each QE,
and $D[\hat{O}]\hat{\rho}=\hat{O}\hat{\rho}\hat{O}^\dag
-\frac{1}{2}\hat{O}^\dag\hat{O}\hat{\rho}-\frac{1}{2}\hat{\rho}\hat{O}^\dag\hat{O}$ for a given operator $\hat{O}$.
In the weak-coupling limit, we can neglect cavity damping terms since $n_\text{ph}\kappa\ll\Gamma$.
In Fig.~7(b), we plot the fidelity as a function of evolution time
and show a high-fidelity preparation of the target state.

%

\end{document}